\documentclass[sigconf,screen,authorversion]{acmart}
\AtBeginDocument{%
  }

\copyrightyear{2026}
\acmYear{2026}
\setcopyright{cc}
\setcctype{by}
\acmConference[CHI EA '26]{Extended Abstracts of the 2026 CHI Conference on Human Factors in Computing Systems}{April 13--17, 2026}{Barcelona, Spain}
\acmBooktitle{Extended Abstracts of the 2026 CHI Conference on Human Factors in Computing Systems (CHI EA '26), April 13--17, 2026, Barcelona, Spain}
\acmDOI{10.1145/3772363.3799295}
\acmISBN{979-8-4007-2281-3/2026/04}

\usepackage{xspace}
\usepackage{subfig}
\usepackage{acmart-taps}

\newcommand{\aisa}{\texttt{AI-SA}\xspace}
\newcommand{\aionly}{\texttt{AI-Only}\xspace}
\newcommand{\gmaps}{\texttt{GMaps}\xspace}

\begin{document}

\pagestyle{plain}
\pagenumbering{arabic}

\title{Navig-AI-tion: Navigation by Contextual AI and Spatial Audio}

\author{Mathias N. Lystb{\ae}k}
\orcid{0000-0001-6624-3732}
\email{mathiasl@cs.au.dk}
\affiliation{%
  \institution{Google}
  \city{Mountain View}
  \country{USA}
}
\affiliation{%
  \institution{Aarhus University}
  \city{Aarhus}
  \country{Denmark}
}

\author{Haley Adams}
\orcid{0000-0002-7329-7840}
\email{haleyaadams@google.com}
\affiliation{%
  \institution{Google}
  \city{Mountain View}
  \country{USA}
}

\author{Ranjith Kagathi Ananda}
\orcid{0009-0006-4047-2987}
\email{ranjithkagathi@google.com}
\affiliation{%
  \institution{Google}
  \city{Mountain View}
  \country{USA}
}

\author{Eric J Gonzalez}
\orcid{0000-0002-2846-7687}
\email{ejgonz@google.com}
\affiliation{%
  \institution{Google}
  \city{Seattle}
  \country{USA}
}

\author{Luca Ballan}
\orcid{0009-0006-7328-9599}
\email{ballanlu@google.com}
\affiliation{%
  \institution{Google}
  \city{Mountain View}
  \country{USA}
}

\author{Qiuxuan Wu}
\orcid{0009-0005-6503-3393}
\email{qiuxuanwu@google.com}
\affiliation{%
  \institution{Google}
  \city{Mountain View}
  \country{USA}
}

\author{Andrea Colac\c{c}o}
\orcid{0009-0001-6661-2216}
\email{andreacolaco@google.com}
\affiliation{%
  \institution{Google}
  \city{Mountain View}
  \country{USA}
}

\author{Peter Tan}
\orcid{0009-0009-2872-184X}
\email{ptan@google.com}
\affiliation{%
  \institution{Google}
  \city{Mountain View}
  \country{USA}
}

\author{Mar Gonzalez-Franco}
\orcid{0000-0001-6165-4495}
\email{margon@google.com}
\affiliation{%
  \institution{Google}
  \city{Seattle}
  \country{USA}
}

\renewcommand{\shortauthors}{Lystb{\ae}k et al.}

\begin{abstract}
Audio-only walking navigation can leave users disoriented, relying on vague cardinal directions and lacking real-time environmental context, leading to frequent errors. To address this, we present a novel system that integrates a Vision Language Model (VLM) with a spatial audio cue. Our system extracts environmental landmarks to anchor navigation instructions and, crucially, provides a directional spatial audio signal when the user faces the wrong direction, indicating the precise turn direction. In a user study (n=12), the spatial audio cue with VLM reduced route deviations compared to both VLM-only and Google Maps (audio-only) baseline systems. Users reported that the spatial audio cue effectively supported orientation and that landmark-anchored instructions provided a better navigation experience over audio-only Google Maps. This work serves as an initial look at the utility of future audio-only navigation systems for incorporating directional cues, especially real-time corrective spatial audio.
\end{abstract}


 \begin{teaserfigure}
   \centering
   \includegraphics[width=\textwidth]{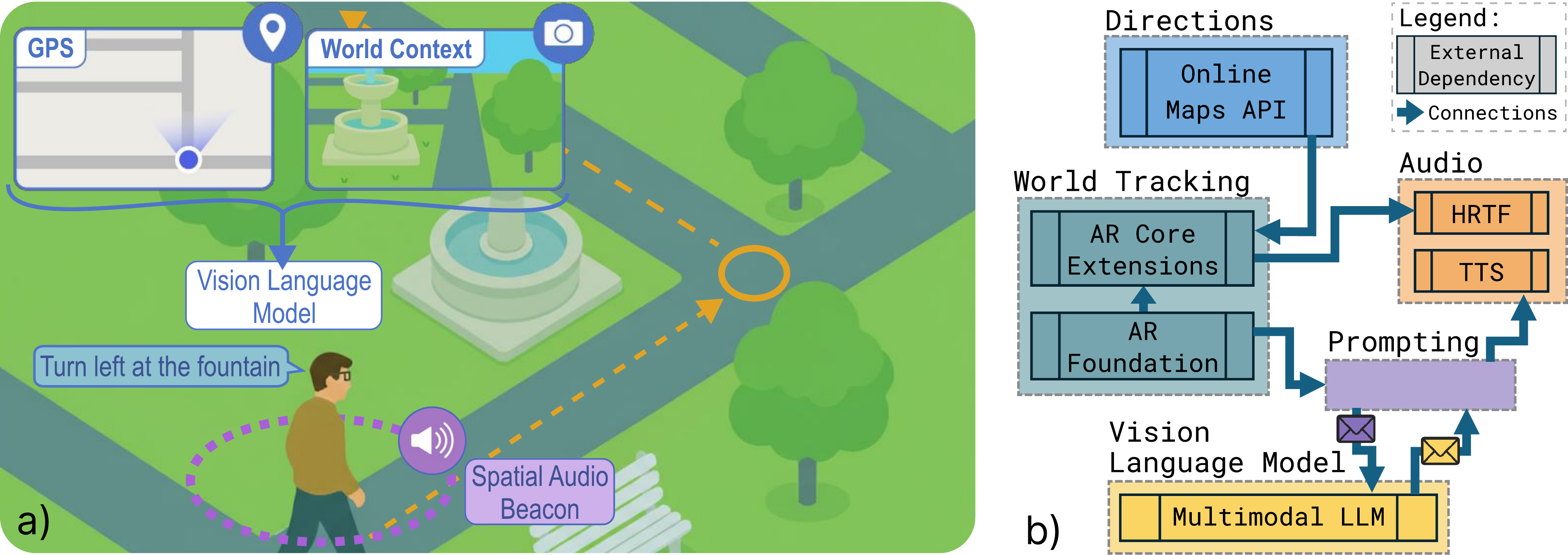}
   \caption{Augmented navigation using AI and spatial audio for display-less smart glasses (a). A simplified overview of the system architecture (b).}
   \Description{Two panels. (a) An illustration of a walking user receiving a Spatial Audio Beacon and a spoken instruction, 'Turn left at the fountain', produced by a Vision Language Model using GPS and world context. (b) A system architecture flow diagram: Online Maps API connects to Directions, which connects to World Tracking; AR Core Extensions and AR Foundation feed World Tracking and Prompting; World Tracking connects to Prompting and Audio; Prompting connects to Audio and Vision Language Model; Vision Language Model connects to a Multimodal LLM in the cloud; Audio connects to a Head-Related Transfer Functions (HRTFs) Audio pipeline and Text-to-Speech (TTS).}
   \label{fig:teaser}
 \end{teaserfigure}

\maketitle

\section{Introduction}
Traditional GPS navigation has enabled a mobility revolution, yet there are legitimate questions about how much engagement with the environment current systems provide~\cite{leshed2008car}. 
Modern turn-by-turn navigation excels at getting us from A to B, but growing evidence suggests that this could come at the cost of understanding and remembering the spaces we navigate~\cite{ishikawa2008wayfinding, gardony2013navigational, dahmani2020habitual}. 
%
Without coupling with the environment, these systems risk reducing navigation to blindly following an endless list of instructions~\cite{gonzalez2021gps}. 

Head-worn devices with cameras, like AR glasses, open new possibilities for enhancing navigation, particularly in complex walking scenarios where minimal screen interactions are desirable.
Prior work has shown that spatial audio cues can provide more seamless orientation guidance, subtly assisting users in re-orienting themselves~\cite{clemenson2021rethinking, berger2018generic, liu2022characterizing, Holland02AudioGPS}. 
%
Meanwhile, head-attached cameras combined with multimodal Large Language Models (LLMs), or Vision Language Models (VLMs), can generate contextual, landmark-based navigation instructions, e.g., replacing ``turn left in 50 feet'' with ``turn left at the fountain'' (cf. \autoref{fig:teaser}a). 

The combination of VLM-powered visual context and spatial audio has the potential to create a less disruptive navigation experience grounded in both sensory input and the environment. 
This is precisely the approach we prototype and evaluate in this paper.


While our goal is to support orientation in audio‑only contexts, our preliminary study did not find improvements in spatial awareness (pointing accuracy). However, we observe reductions in route deviations and distance walked.

\section{Related Work}


The pervasive use of turn-by-turn navigation apps (e.g., Google Maps, Apple Maps, Waze) is a testament to their effectiveness for reaching destinations~\cite{miola2024gps}. 
However, this convenience can come at a cognitive cost: 
the more people rely on GPS navigation, the worse their perceived navigation skills and  environmental knowledge tends to become~\cite{topete2024gps, miola2024development}.

\subsection{Audio cues for navigation}
Audio cues have been shown to effectively support walking navigation as an alternative to screen-based instructions~\cite{clemenson2021rethinking, Spagnol18UseOfSpatialAudio, Holland02AudioGPS, McGookin09AudioBubbles, Warren05NavigationMusic, Strachan05GPSTunes}, with spatial audio in particular helping users find the correct direction~\cite{clemenson2021rethinking, Simpson05SpatialAudioNavigationAid, Spagnol18UseOfSpatialAudio}. 
Notably, ~\citeauthor{clemenson2021rethinking} introduced Soundscape~\cite{clemenson2021rethinking}, which guided users via continuous spatial audio and improved spatial recollection of points of interest (POIs). 
Our approach is similar, however, we only present the spatial audio cue when the user is not facing the suggested direction, lowering auditory load.

\subsection{VLMs for navigation}
Recent advances in LLMs enable richer, more context-aware descriptions of surroundings and more natural interactions~\cite{abreu2024parse}, especially multimodal variants (VLMs) that integrate visual and textual data.
To the best of our knowledge, few systems integrate VLMs for wayfinding. 
One example is NaviGPT~\cite{zhang2025enhancing}, which combines Apple Maps (route guidance), LiDAR (obstacle detection), and GPT-4 (information processing and instructions) for real-time navigation on iPhone. 
In contrast to NaviGPT, our system is designed exclusively for audio-based delivery and supports display-less devices, anticipating future hardware form factors (e.g., Meta Ray-Bans). 
Additionally, we present a user evaluation yielding key design insights into VLM integration for wayfinding that can guide future development in this emerging field.











\section{System}
The system augments navigation instructions with context about the user and their environment, including orientation and a point-of-view photo. Furthermore, the system provides spatial auditory feedback when the user is not facing the next turn, giving the user an indication of the direction that they should walk. The spatial audio cue is otherwise turned off, allowing the user to listen to their own audio while walking (e.g., music, videos, or podcasts).


The system consists of five components working in sequence to (1) collect navigation instructions, (2) track the user's position and orientation, (3) present audio feedback, (4) capture the user's perspective and prompt the VLM, and (5) extract landmarks and augment the original instruction. See \autoref{fig:teaser}b for a diagram overview.

\subsection{Directions}
Walking directions are queried based on a list of POI GPS coordinates (in this work, we used the Google Maps Platform API),
returning JSON-formatted steps with walking instructions for the desired origin and destination.


\subsection{World Tracking}
The World Tracking component tracks the user and keeps track of where turn waypoints and POIs are relative to the user, also checking if the user has reached a turn.
The system only captures an image from the user's point-of-view when it is about to prompt the VLM.
%
%
AR Foundation provides the core functionality for phone tracking, camera capture, and spatial audio positioning based on user orientation.
%
AR Core Extensions adds accurate GPS coordinate capture and virtual anchor placement.
%
User orientation is determined via Google's Visual Positioning System (VPS). 
%
\aptLtoX[graphic=no, type=html]{
}{
\begin{figure*}
    \centering
    \subfloat[Routes.]{%
        \includegraphics[height=125pt]{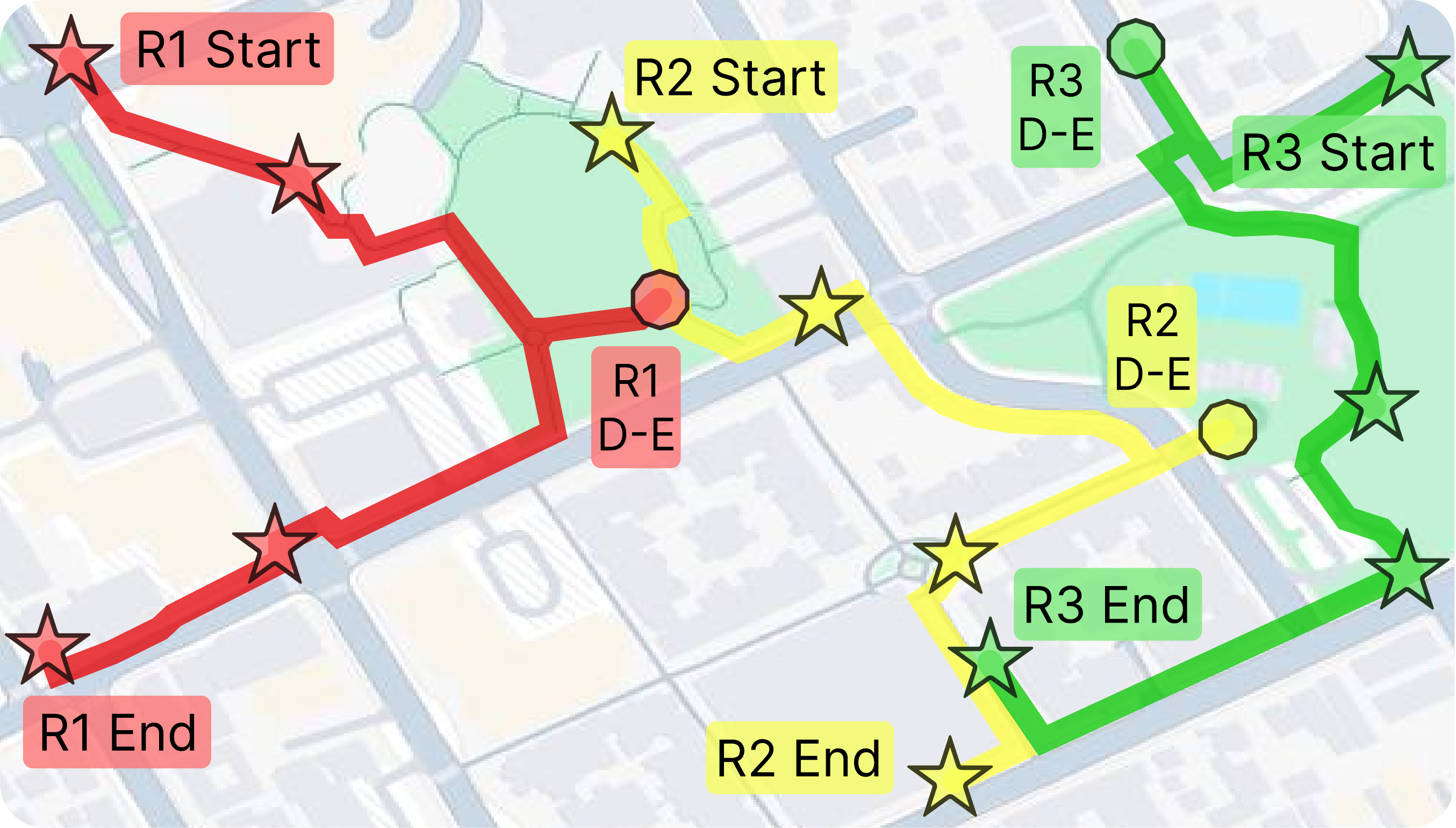}\label{subfig:routes}%
    }
    \hfill
    \subfloat[Apparatus.]{%
        \includegraphics[height=125pt]{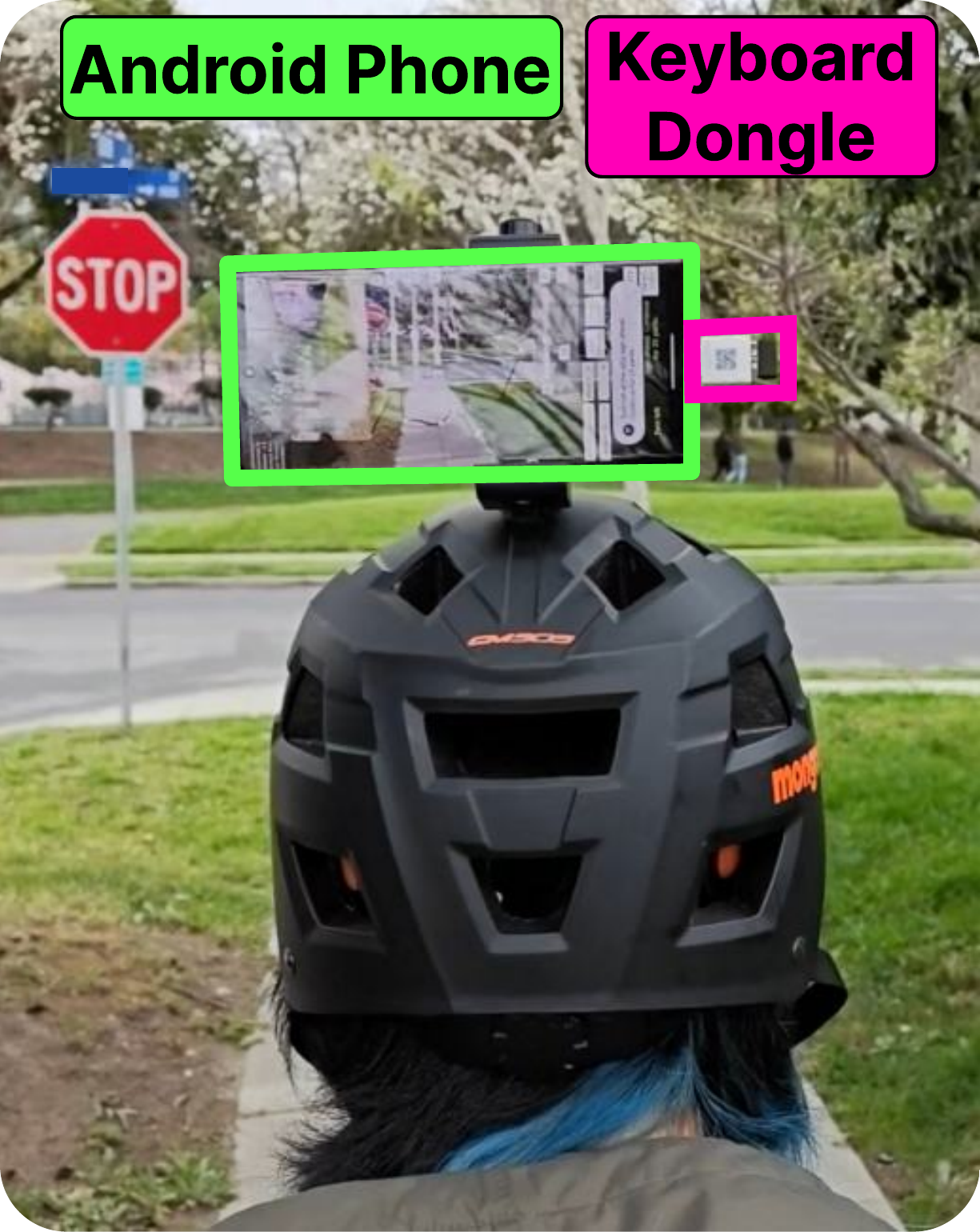}\label{subfig:apparatus}%
    }
    \hfill
    \subfloat[Results on Distance and Deviations.]{%
        \includegraphics[height=125pt]{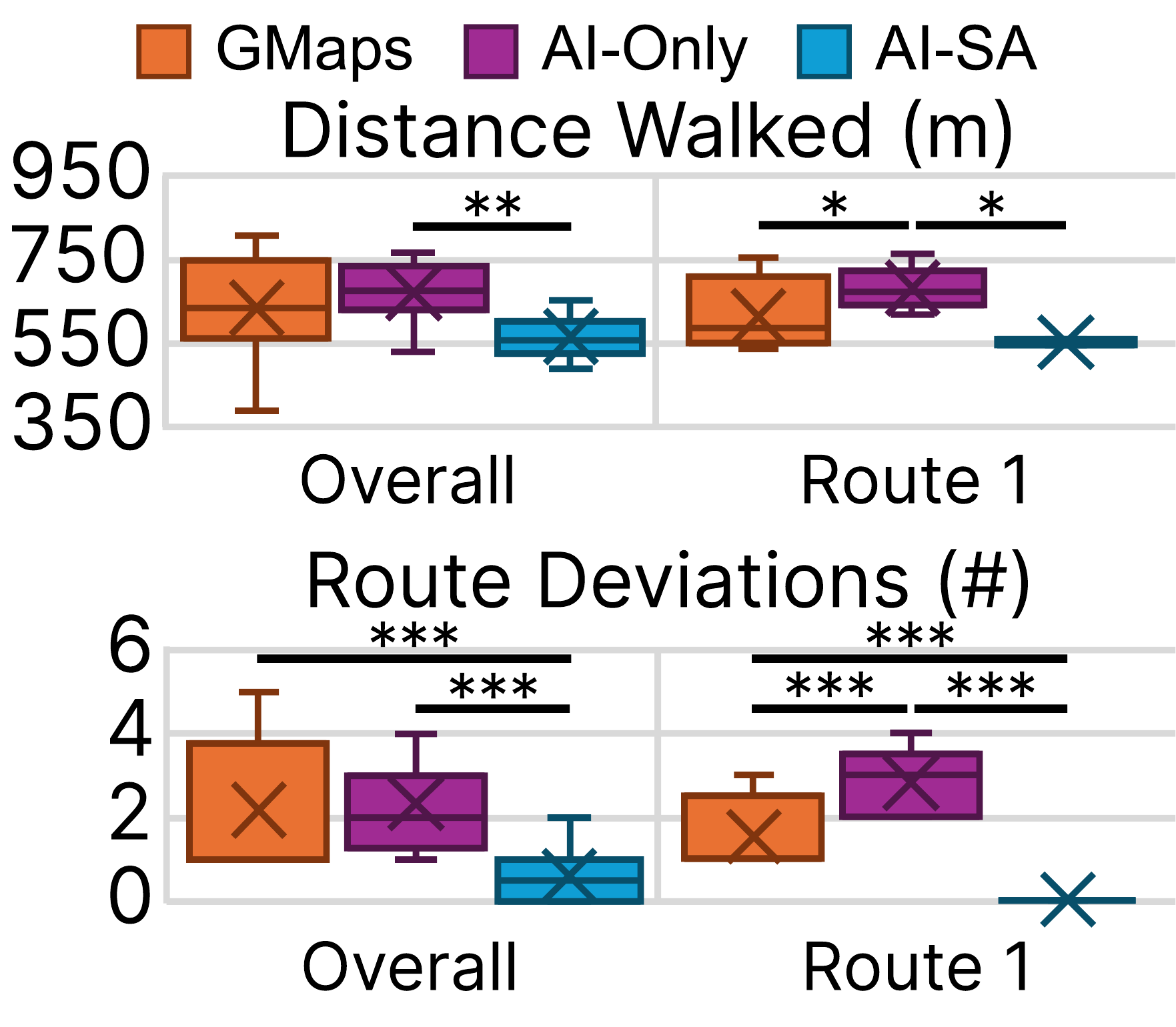}\label{subfig:results}%
    }
    \caption{(a) Three color-coded routes, marked with route start and end for each route RX, where ``X'' is 1 to 3, along with color-matched stars and circles indicating landmarks along the routes. ``RX D-E'' indicates the dead end for each route. (b) The apparatus setup for the prototype and evaluation. (c) Box plots showing results on Distance Walked (m) and Route Deviations (count). Whiskers indicate minimum and maximum. Statistical significance is shown as * for $p<0.05$, ** for $p<0.01$, and *** for $p<0.001$.}
    \Description{Three subfigures. (a) A map with three color-coded walking routes labeled R1 (red), R2 (yellow), and R3 (green). Each route shows 'Start,' 'D-E' (dead end), and 'End,' with matching colored star markers along the paths. (b) Photo of the study apparatus seen from behind: a helmet-mounted setup with two labeled boxes, 'Android Phone' (green label) and 'Keyboard Dongle' (magenta label). (c) Two small box-plot charts comparing conditions 'GMaps,' 'AI-Only,' and 'AI-SA.' The upper chart is titled 'Distance Walked (meters)' with tick marks from roughly 150 to 950. The lower chart is titled 'Route Deviations (count).' Each chart has groups for 'Overall' and 'Route 1,' whiskers indicating minimum and maximum, and significance bars marked with *, **, or ***.}
    \label{fig:routes-results}
\end{figure*}
}

\subsection{Audio}
The audio component controls spatial audio cue playback and Text-To-Speech (TTS) feedback (Android TTS API running on-device). 
To accurately simulate real-world audio characteristics, we utilize Head-Related Transfer Functions (HRTFs), simulating how sound passes through human ears based on the relative position of the audio cue and human ear topology (in this work, we use the Steam Audio SDK).
Similar to~\cite{clemenson2021rethinking}, the spatial audio source is placed $1$ meter from the user (cf. \autoref{fig:teaser}a), conveying turn direction, but not distance.
%
%
The spatial audio cue is presented as a repeating ``waterfall'' sound, which we found suitable for individuals to accurately locate (validated through a training phase).
As mentioned, the spatial audio cue is not presented to the user at all times, activating when the user’s facing direction deviates by more than $25^\circ$ off from the current turn waypoint (either to the left or the right) and stops playing once the user is facing the current turn waypoint within $25^\circ$. 

\subsection{Prompting}
The prompting component serves as an intermediary between the World Tracking, Audio, and VLM components. It captures the context (including camera image, position, orientation, direction and distance to the next turn, and information about the current navigation step), which is sent to the VLM component. Note that the model is initialized with system instructions to steer the model.
Prompting occurs in two scenarios: (1) if the user is stationary while looking around and at some point looks in the direction of the current turn waypoint (within $25^{\circ}$), or (2) when walking close to the turn waypoint (within 30 meters).
When a response is received from the VLM, the contextual navigation instruction is provided to the user via TTS through the Audio component.

\subsection{Vision Language Model}
The VLM component describes the structure of the system instructions provided when creating an instance of the VLM (in this work, we use Gemini 2.0 Flash Exp). The system instructions tell the model what role it should ``play'', provides brief information about the system concept and task description, and the expected input format of an image and metadata.
As a brief overview, the VLM is instructed to:
    Provide egocentric instructions instead of cardinal instructions.
    Detect potential landmarks and include the landmarks' position relative to the user in the instruction, along with detecting a bounding box to mitigate left/right reversal hallucinations, i.e., landmarks on the left mentioned as being on the right..
    If the user is not facing the correct direction, the VLM is instructed to tell the user to turn and position the landmark on their side or behind them.
    If the distance to the next turn is too far ($> 60$ meters), then the VLM is instructed to tell the user to continue forward.
    To make the user aware of their next turn more quickly, the VLM is instructed to present the turn guidance first, followed by the contextual guidance.
    Furthermore, the VLM is explicitly instructed to not hallucinate.
On each prompt, the model parses based on the system instruction, image, and metadata, and outputs the most likely response. 


\aptLtoX[graphic=no, type=html]{
\begin{figure*}
    \centering
    \subfloat[Routes.]{%
        \includegraphics[height=125pt]{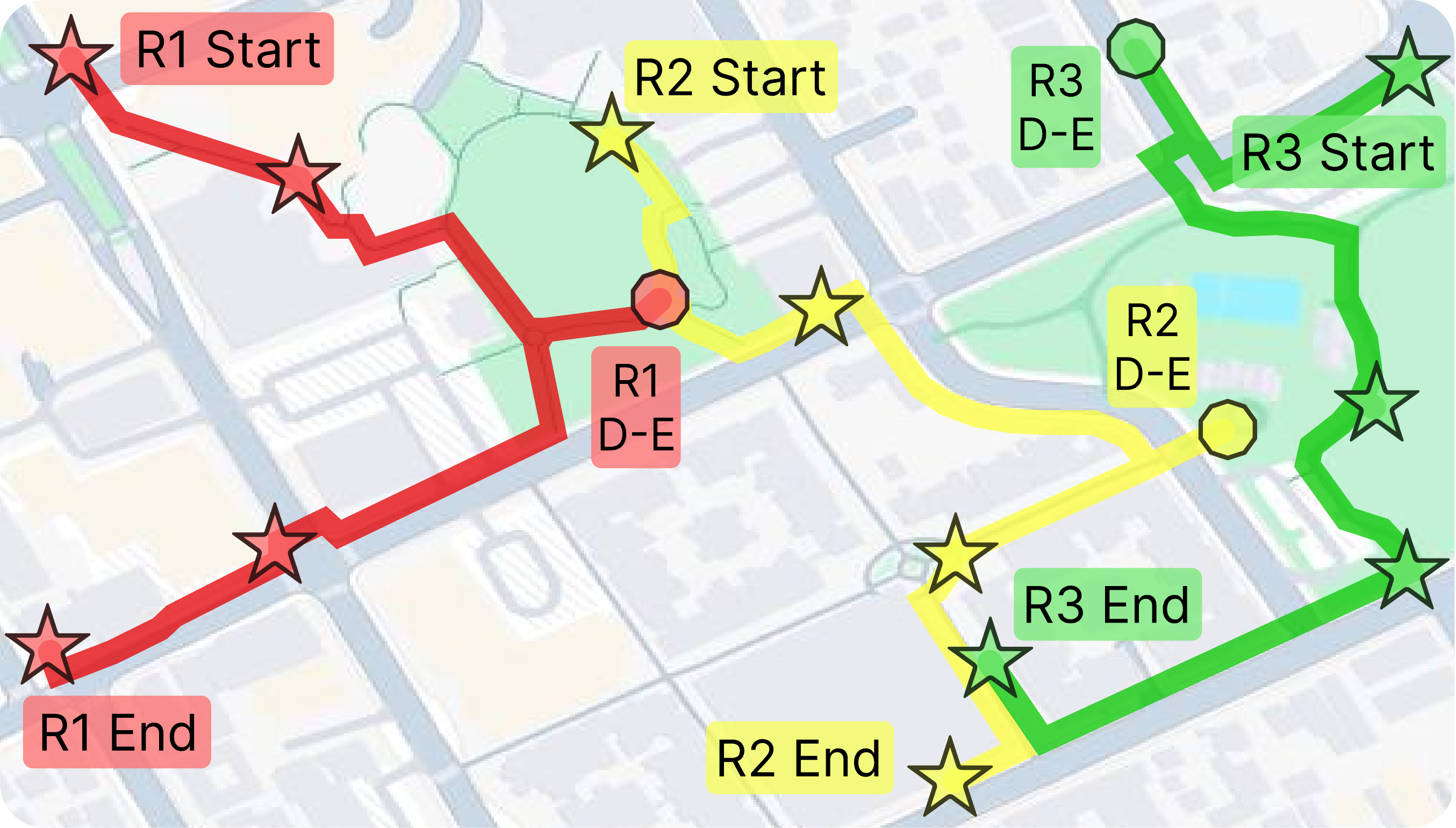}\label{subfig:routes}%
    }
    \hfill
    \subfloat[Apparatus.]{%
        \includegraphics[height=125pt]{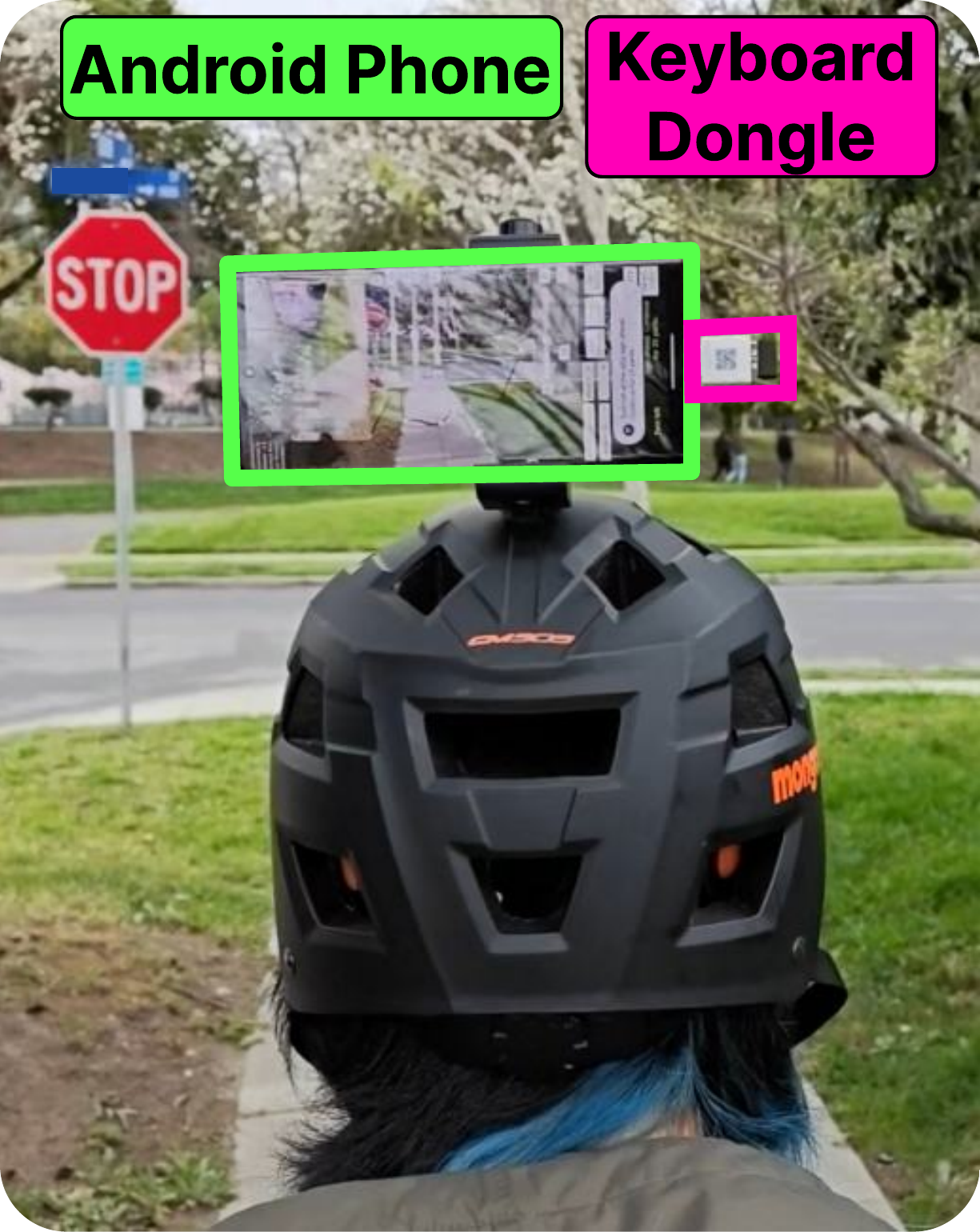}\label{subfig:apparatus}%
    }
    \hfill
    \subfloat[Results on Distance and Deviations.]{%
        \includegraphics[height=125pt]{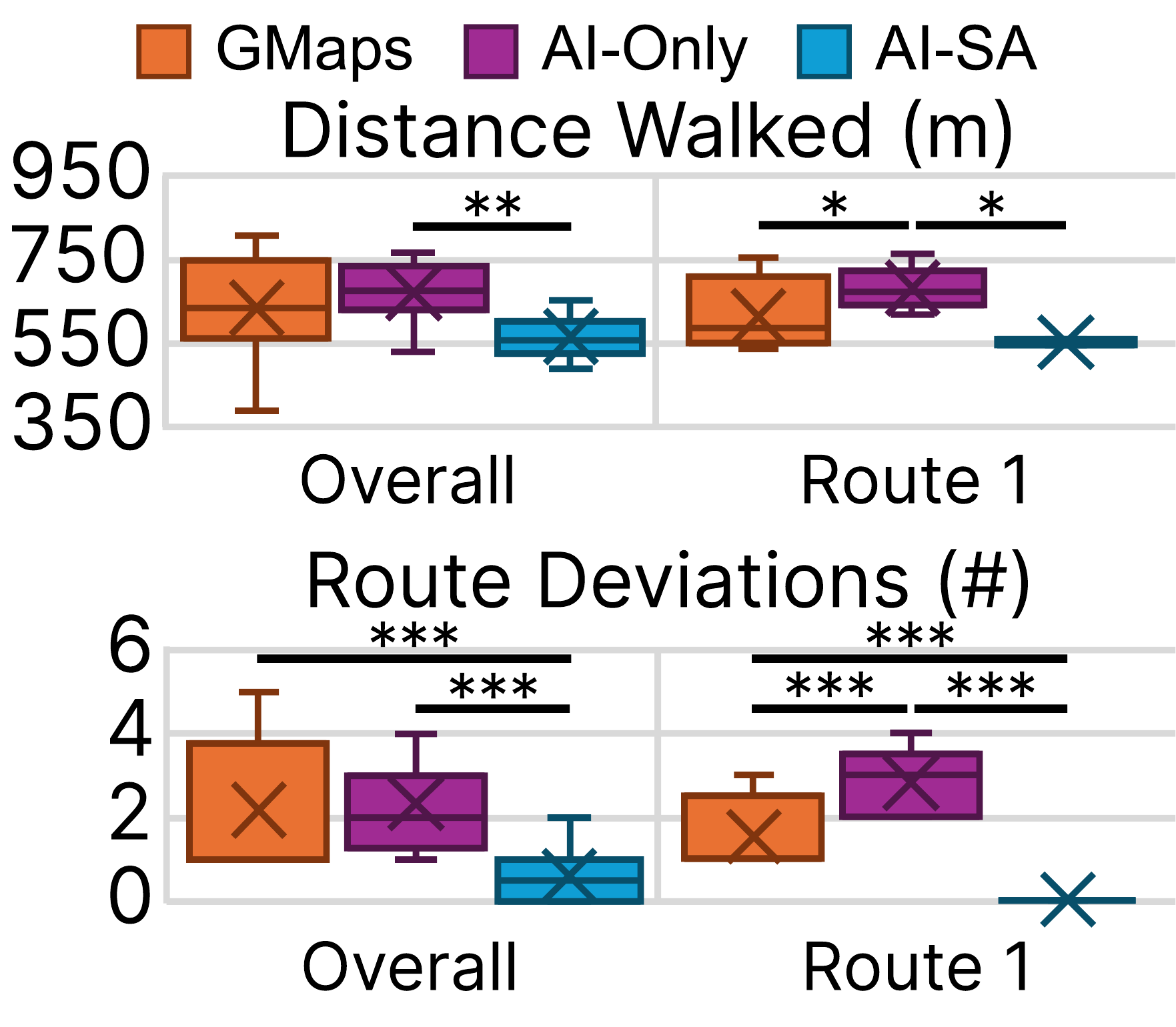}\label{subfig:results}%
    }
    \caption{(a) Three color-coded routes, marked with route start and end for each route RX, where ``X'' is 1 to 3, along with color-matched stars and circles indicating landmarks along the routes. ``RX D-E'' indicates the dead end for each route. (b) The apparatus setup for the prototype and evaluation. (c) Box plots showing results on Distance Walked (m) and Route Deviations (count). Whiskers indicate minimum and maximum. Statistical significance is shown as * for $p<0.05$, ** for $p<0.01$, and *** for $p<0.001$.}
    \Description{Three subfigures. (a) A map with three color-coded walking routes labeled R1 (red), R2 (yellow), and R3 (green). Each route shows "Start," "D-E" (dead end), and "End," with matching colored star markers along the paths. (b) Photo of the study apparatus seen from behind: a helmet-mounted setup with two labeled boxes, "Android Phone" (green label) and "Keyboard Dongle" (magenta label). (c) Two small box-plot charts comparing conditions "GMaps," "AI-Only," and "AI-SA." The upper chart is titled "Distance Walked (meters)" with tick marks from roughly 150 to 950. The lower chart is titled "Route Deviations (count)." Each chart has groups for "Overall" and "Route 1," whiskers indicating minimum and maximum, and significance bars marked with *, **, or ***.}
    \label{fig:routes-results}
\end{figure*}
}{
}

\section{Preliminary Evaluation}
This study serves as a preliminary evaluation, aiming to gather an initial understanding of user behavior and experience with our contextual AI and spatial audio system. Participants walked predefined routes (cf. \autoref{subfig:routes}) while following navigation instructions and audio cues.
We compare our system, with VLM instruction augmentation and spatial audio (\aisa), against a version with just VLM instruction augmentation (\aionly), and a state-of-the-art baseline in the form of Google Maps (audio-only) with detailed walking navigation enabled, (\gmaps).

The study used a within-subject design with one independent variable \textit{condition} (\aisa, \aionly, and \gmaps) in counterbalanced order to limit order effects. 
To eliminate route learning, we prepared three different routes, although in the same area, which were also counterbalanced. 

The three routes were similar in distance (550-650 meters), estimated time (8-9 minutes), included the same number of POIs (5), and a single ``dead end'' each (see \autoref{tab:route-comparison} for an overview). The dead end required  participants to return along the path they came from, exposing a core challenge when routes are not ``linear'' and include backtracking. 


To capture the participants' point-of-view as best as possible, an Android phone was mounted to a helmet (cf. \autoref{subfig:apparatus}), simulating the head-worn device form factor, which was especially important for the spatial audio, as the orientation of the phone determines how the spatial audio updates.


Participants were tasked to walk to each of the POIs along the routes and remember the POIs along the route for a pointing task in the end.
Notably, participants were not informed about the inclusion of dead ends.
Once ready and the study had started, they were told to walk in the direction they believed they should.
The experiment conductor walked closely behind the participant to monitor the application, ensuring that the study went smoothly. Whenever the participants walked too far off the route ($> 10$ meters), the conductor intervened, noted the deviation, and told the participant that they walked in the wrong direction. The participant would then try another direction.
Upon reaching the end of the route, the participants pointed with their head towards the beginning POI, then subsequent POIs along the route, with each angular offset being logged to gauge differences in spatial awareness.

\subsection{Results}

We recruited $12$ participants ($6$ male, $5$ female, $1$ non-binary), consisting mainly of software engineers. Participants' ages ranged from 18 to 55. All participants had normal, or corrected-to-normal, vision.

ANOVA was used to analyze normally distributed quantitative data (with Greenhouse-Geisser correction), while Friedman tests (with post-hoc Conover tests) were used for non-normally distributed data (all Holm-Bonferroni corrected).
\autoref{subfig:results} shows only significant results for brevity, see 
\autoref{appendix:results}
in supplementary materials for additional graphs and user feedback.
Route-level analyzes use only the subset of participants assigned to that route under the counterbalanced design, resulting in smaller degrees of freedom. See 
\autoref{appendix:gps-paths}
for all recorded paths annotated with deviations.
\textbf{Distance Walked}, measured by GPS ($F(2,22)=5.017, p=0.016, \eta^2=0.313$), shows that our participants overall walked shorter paths with \aisa than \aionly ($p=0.004$), indicating that they followed the paths more accurately. 
Particularly on Route 1 ($F(2,4)=40.959, p=0.002, \eta^2=0.953$), we found that participants walked longer paths with \aionly compared to both \gmaps ($p=0.02$) and \aisa ($p=0.048$). 
Participants overall had fewer \textbf{Route Deviations} ($> 10m$ off-route) ($\chi^2=14.55, p<0.001, W=0.606$) with \aisa than the other conditions (both $p<0.001$). 
On Route 1 ($\chi^2=6, p=0.05, W=1$), \aionly exhibited more route deviations than both \gmaps ($p<0.001$) and \aisa ($p<0.001$), while \gmaps exhibited more route deviations than \aisa ($p<0.001$). 
\textbf{Pointing Accuracy} was measured to gauge spatial awareness, as the angular error between the participant's pointing direction and the actual direction of the four POIs from the last POI.
We found no differences in Pointing Accuracy overall or any of the routes. 
Prompt-to-response latency was measured at $3.31 s$ on average ($\pm 0.81 s$, from $1.48 s$ to $11.29 s$). 

Participants generally rated \aisa highest (only once rating the condition 2nd), while \aionly (rated 3rd three times) was generally rated higher than \gmaps (rated 3rd 9 times).
Several participants described the spatial audio cue as helpful for verifying if they were on the right track and for orienting themselves. 
%
The continuous nature of the spatial audio cue was appreciated by some, making it easier to follow. 
%
However, some found the spatial audio cue could be drowned out by other sounds or become easy to ignore if not focused. 
%
%
%
Participants mentioned appreciating specific and notable landmarks
as they were easy to see and provided clear reference points. 
%
However, more common landmarks
were often found confusing and unhelpful,
making it difficult to identify the correct one. 
%
%
%
%
The use of cardinal directions in \gmaps was widely perceived as unhelpful, especially without a map or compass at hand. Participants often did not know which direction was north, south, east, or west, leading them to guess.

%



\section{Discussion \& Future Work}

While the addition of landmarks to navigation instructions has been shown to improve user performance and experience~\cite{zhang2025enhancing}, our results are initially not better than the state-of-the-art \gmaps, other than route deviations; we saw no differences in pointing accuracy. 
However, users notably preferred the VLM-augmented instructions (\aionly and \aisa) over \gmaps, mostly due to the inclusion of landmarks and conversion of cardinal directions to egocentric directions, while especially the spatial audio cue of \aisa rated highly with participants.

\subsection{Route dissimilarities}
Despite our efforts to keep the three routes comparable, Route 1 appears more complex, particularly in the \aionly condition. Post-hoc comparisons, summarized in \autoref{tab:route-comparison} (inspired by~\cite{Raubal1998ComplexityWayfinding}), reveal that while Route 1 had fewer intersections and turns, it had a higher average number of alternative paths per intersection. The spatial audio cue inherently disambiguates such choices, whereas turn-by-turn conditions may require participants to think and make assumptions, potentially leading to more deviations. 
As shown in \autoref{appendix:gps-paths},
the dead-end on Route 1 caused notably more deviations in \aionly than in the other conditions (dead-end deviations: $\aionly=8$, $\gmaps=3$, $\aisa=0$), likely due to its placement at an intersection (see \autoref{subfig:routes}), which left participants uncertain about which way to turn.
\begin{table}
    \centering
    \caption{An overview of route characteristics. $\pm$ indicates standard deviation. Note that the 180\textdegree{} turn at the dead end is excluded from the mean turn angle.}
    \small
    \begin{tabular}{p{0.25\columnwidth}|p{0.2\columnwidth}|p{0.2\columnwidth}|p{0.2\columnwidth}}
        Measure & Route 1 & Route 2 & Route 3 \\ 
        \hline\hline
        Distance & 650 & 550 & 600 \\ \hline
        Estimated time & 9 min. & 8 min. & 9 min. \\ \hline
        \# of intersections & 9 & 10 & 10 \\ \hline
        Mean alternative paths & $2.78$ $(\pm 0.83)$ & $2.5$ $(\pm 0.53)$ & $2.4$ $(\pm 0.7)$ \\ \hline
        Mean intersection turn angle & $67.28^\circ$ $(\pm 28.95^\circ)$ & $64.97^\circ$ $(\pm 23.05^\circ)$ & $80.65^\circ$ $(\pm 12.76^\circ)$ \\ \hline
        Turns & 8 & 9 & 10
    \end{tabular}
    \label{tab:route-comparison}
    \Description{Table 1 summarizes characteristics of three routes used in the study. Columns are Route 1, Route 2, and Route 3. Rows list distance (meters), estimated time (minutes), number of intersections, mean alternative paths per intersection, mean intersection turn angle in degrees with standard deviation, and number of turns. The 180-degree turn at the dead end is excluded from the mean turn angle. Route 1: Distance 650 meters; estimated time 9 min; 9 intersections; mean alternative paths 2.78 (plus or minus 0.83); mean intersection turn angle 67.28 degrees (plus or minus 28.95 degrees); 8 turns. Route 2: Distance 550 meters; estimated time 8 min; 10 intersections; mean alternative paths 2.5 (plus or minus 0.53); mean intersection turn angle 64.97 degrees (plus or minus 23.05 degrees); 9 turns. Route 3: Distance 600 meters; estimated time 9 min; 10 intersections; mean alternative paths 2.4 (plus or minus 0.7); mean intersection turn angle 80.65 degrees (plus or minus 12.76 degrees); 10 turns.}
\end{table}

\subsection{Spatial Sound}
Our results highlight that the spatial audio cue was critical to keeping users on their path,
while the VLM-augmented instructions served more as preemptive information to prepare participants for an upcoming turn.
Rather than relying on cardinal or egocentric directions, the spatial audio gives a clear, continuously updating signal for where the next turn is.
We therefore believe spatial audio would play a critical role in display-less scenarios, such as when the phone is pocketed or when wearing smart glasses.




\subsection{Choice of landmarks}
Participants noted a preference for more unique landmarks, as simply being told to ``turn left at the tree on your right'' is unhelpful, especially in a park with many trees.
We believe that these cases could be mitigated through prompt engineering and task-specific fine-tuning.
One participant mentioned wanting more consistency in landmark choice when looking in the same area, which could potentially be achieved by sending a history of recent prompts and responses to the model.

\subsection{Hallucinations} 
Although we instructed the VLM to avoid hallucinations, others have found them inherent to LLMs~\cite{Xu2025HallucinationInevitable}. 
Despite attempts to correct left/right reversals via bounding box detection, such errors still occurred, causing participant confusion, as did occasional object misidentification.

\subsection{Prompt latency}
The average latency of $\approx$3.3s is largely attributable to Gemini 2.0 Flash prioritizing speed over quality. At an average walking speed of $\approx$1.38m/s~\cite{Bohannon11WalkingSpeed}, this equates to $\approx$4.5m of travel, meaning landmarks closer than $\approx$5m risk causing confusion rather than helping. While manageable on foot, the issue scales significantly at higher speeds: $\approx$18.3m at 20km/h (biking) and $\approx$45.8m at 50km/h (city driving), making latency reduction critical for broader use cases.

\subsection{Street View Augmentation}
The quality of images captured from the user's perspective greatly affects what the VLM can extract.
Existing data from tools like Google Street View could partially address this~\cite{Chen19Touchdown, Hermann20StreetNav, Mirowski19StreetLearn}, enabling landmark extraction at route initialization and reducing latency, while also reducing battery consumption for camera-less devices. Prompts could be adapted to favor static landmarks, such as large buildings.
However, since Street View may be outdated, landmarks could become unavailable or occluded from the user's view. Perhaps a hybrid approach could find landmarks online and verify validity in real time.
%
%


\subsection{Variants of hardware degradation}
Although our system requires stereo audio, it could be adapted for reduced functionality to broaden accessibility, including mono audio devices.
The core parts needed are a navigation directions source (e.g., Google Maps or OpenStreetMap), a way of capturing the environment (be it in real time with a camera or online through Street View or similar), a low-latency VLM, and text-to-speech.
Correct or incorrect facing directions could also be conveyed via auditory icons.

\subsection{\textbf{Limitations}}
As this is a small exploratory study ($n=12$), per-route comparisons rely on small subsets and should be interpreted with caution. The routes could have been more similar, as significant differences between conditions were only indicated on Route 1, though we believe varied layouts are important for capturing real-world variability.

The model used, Gemini 2.0 Flash Exp, has since been deprecated, which limits interpretability. However, we expect newer model versions to perform at least as well given architectural improvements.

Finally, while we report prompt-to-response latency, component-level breakdowns could have offered additional insight, though the prompt-to-response measure captures the major portion of overall latency.

\section{Conclusion}
In this work, we introduced a novel audio-only navigation system that leverages spatial audio and contextual awareness to improve the user experience. Our findings indicate that the integration of particularly spatial audio cues (with VLM-generated landmark-anchored instructions) leads to improved navigation efficiency compared to VLM-only systems and traditional navigation apps like Google Maps (audio-only), although no improvement to spatial awareness was found.
Our users preferred egocentric directions, from the VLM, over cardinal directions, from Google Maps. Similarly, users noted that the landmark-anchored instructions were easier to follow, although mainly when the landmark is unique.
This work indicates benefits to be had from utilizing directional and environmental context to better guide users without access to a screen.

%

\bibliographystyle{ACM-Reference-Format}
\bibliography{main}

\clearpage
\appendix

\section{Additional Results}\label{appendix:results}
This appendix contains additional results that could not make it into the main paper, including additional figures and user feedback.

\subsection{Additional quantitative figures}
\begin{figure*}[ht]
    \centering
    \includegraphics[width=.7\textwidth]{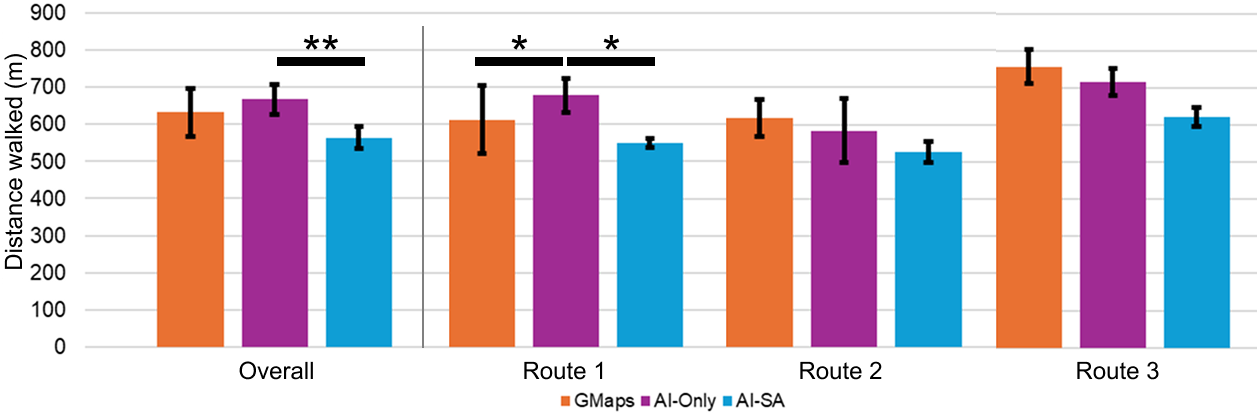}
    \caption{Results on the Distance Walked overall and for each route separately.}
    \Description{A bar graph comparing the average distance walked in meters for three conditions: AI-SA, AI-Only, and GMaps. The graph shows that, on average, AI-SA has the shortest distance walked, followed by GMaps, and then AI-Only.}
    \label{fig:res-distance}
\end{figure*}

\begin{figure*}[ht]
    \centering
    \includegraphics[width=.7\textwidth]{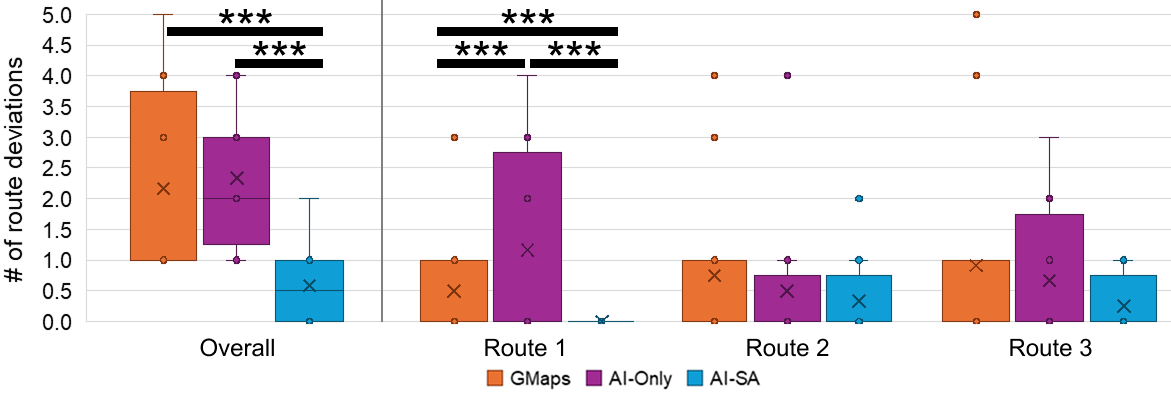}
    \caption{Results on the Number of Deviations overall and for each route separately.}
    \Description{A box plot graph comparing the average number of route deviations for three conditions: AI-SA, AI-Only, and GMaps. The graph shows that AI-SA has the fewest deviations, followed by AI-Only, and then GMaps.}
    \label{fig:res-num-deviations}
\end{figure*}

\begin{figure*}[ht]
    \centering
    \includegraphics[width=.7\textwidth]{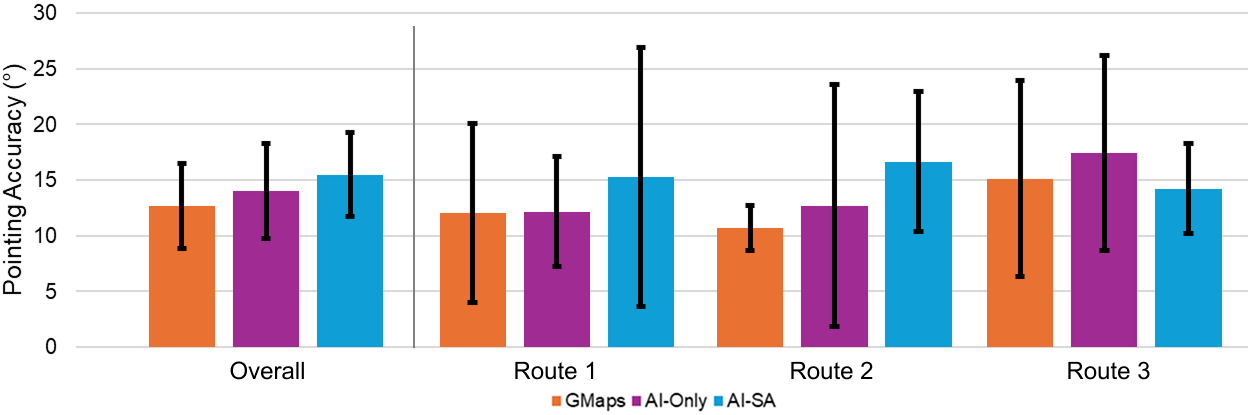}
    \caption{Results on Pointing Accuracy overall and for each route separately.}
    \Description{A bar graph comparing the average pointing error in degrees for three conditions: AI-SA, AI-Only, and GMaps. The graph shows that, on average, GMaps has the lowest pointing error, followed by AI-Only, and then AI-SA.}
    \label{fig:res-pa}
\end{figure*}

\begin{figure*}[ht]
    \centering
    \includegraphics[width=.35\textwidth]{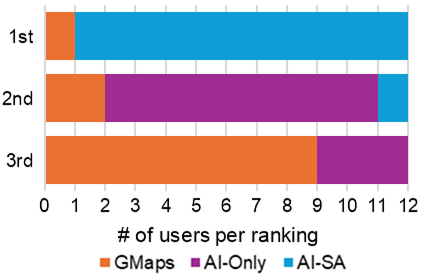}
    \caption{Results on user preference/rankings of the three conditions.}
    \Description{A graph showing the distribution of condition rankings by participants. The x-axis lists the condition ranking (1st, 2nd, 3rd), and the y-axis shows the number of participants. The graph indicates that AI-SA was most frequently ranked 1st, followed by AI-Only, with GMaps being least preferred.}
    \label{fig:res-rankings}
\end{figure*}

\autoref{fig:res-distance} shows the results for Walked Distance across all routes.
\autoref{fig:res-num-deviations} shows the results for Route Deviations across all routes. 
\autoref{fig:res-pa} shows the results on Pointing Accuracy across all routes.
\autoref{fig:res-rankings} shows the results for Ranking.

\subsection{Additional participant feedback}
The following contains additional, and more detailed, participant feedback.
Several participants described the spatial audio cue as helpful for verifying if they were on the right track and for orienting themselves. P2 noted that the ``[audio cue] is clear about direction. Whenever lost track then can listen for the audio'', and P4 stated it ``helped verify that I was pointing in the right direction''. P11 mentioned that it provided ``almost constant feedback'' on whether they were headed the right way.
The continuous nature of the spatial audio cue was appreciated by some, making it easier to follow. P5 noted they rely more on the spatial audio cue because it is faster, and P6 described it as ``more accurate, more continuous''.
However, some found the spatial audio cue could be drowned out by other sounds or become easy to ignore if not focused. P8 preferred a ``louder part of the clip'' so it was clearer, and P12 likened it to ``elevator music'' that could be ``zoned out''.
P7 mentioned needing to get used to the sound initially.
P9 noted that the spatial audio cue helped with direction, while spoken instructions allowed for more fluid movement as it helped them prepare.

Participants mentioned appreciating specific and notable landmarks like a ``sign'' (P12), ``arches'' (P1), ``Wells Fargo'' (P12), ``stairs'' (P8), or a ``silver car'' (P12) as they were easy to see and provided clear reference points. Instructions referencing nearby items like a ``trash can'' were also clear (P5 and P7).
However, more common objects/landmarks such as ``tree'' (P2, P5, P6, P9, P10), ``fire hydrant'' (P1), ``white building'' (P1), or ``lamppost'' (P4, P6) were often found confusing and unhelpful because there could be multiple such objects nearby, making it difficult to identify the correct one. Some participants walked past landmarks without noticing them (P12).
Participants preferred prioritizing unique landmarks over common ones (P6, P10).
Instructions telling participants to turn towards a specific street were considered helpful when street signs were visible (P7, P12).
However, participants mentioned problems occurring when one ``does not know the neighborhood'' (P11) or ``could not see street signs'' (P9), making these instructions less helpful.

The use of cardinal directions in \gmaps was widely perceived as unhelpful (P1, P2, P5, P6, P8, P10, P11, P12), especially without a map or compass at hand (P10, P11). Participants often did not know which direction north, south, east, or west was, leading them to guess.

Participants had mixed opinions about distance estimates. Some found instructions like ``next turn is in how many meters'' to be helpful (P5, \gmaps). Distance affirmation was also appreciated (P2, P7).
However, many participants stated they were not good at judging distances and sometimes felt they had to turn sooner than the indicated distance (P4, P8). P10 mentioned that ``If I can be told that you'll pass two intersections along the way turn right after the third intersection or third block or whatever that that is much more useful than the distance.''

\section{GPS Paths}\label{appendix:gps-paths}
This appendix provides a visual overview of the paths that our participants walked along with annotations for deviations.

\begin{figure*}[ht]
    \centering
    \includegraphics[width=.9\textwidth]{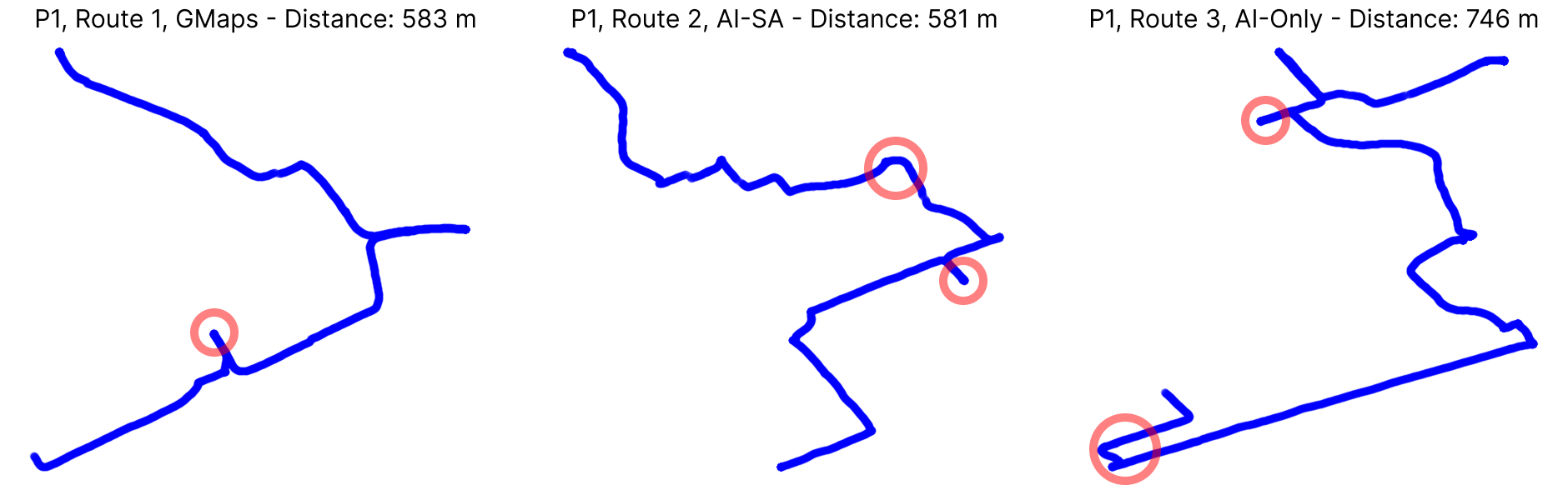}
    \caption{P1's walked paths with 1 deviation in Route 1, 2 in Route 2, and 2 in Route 3.}
    \label{fig:appendix-p1-paths}
\end{figure*}

\begin{figure*}[ht]
    \centering
    \includegraphics[width=.9\textwidth]{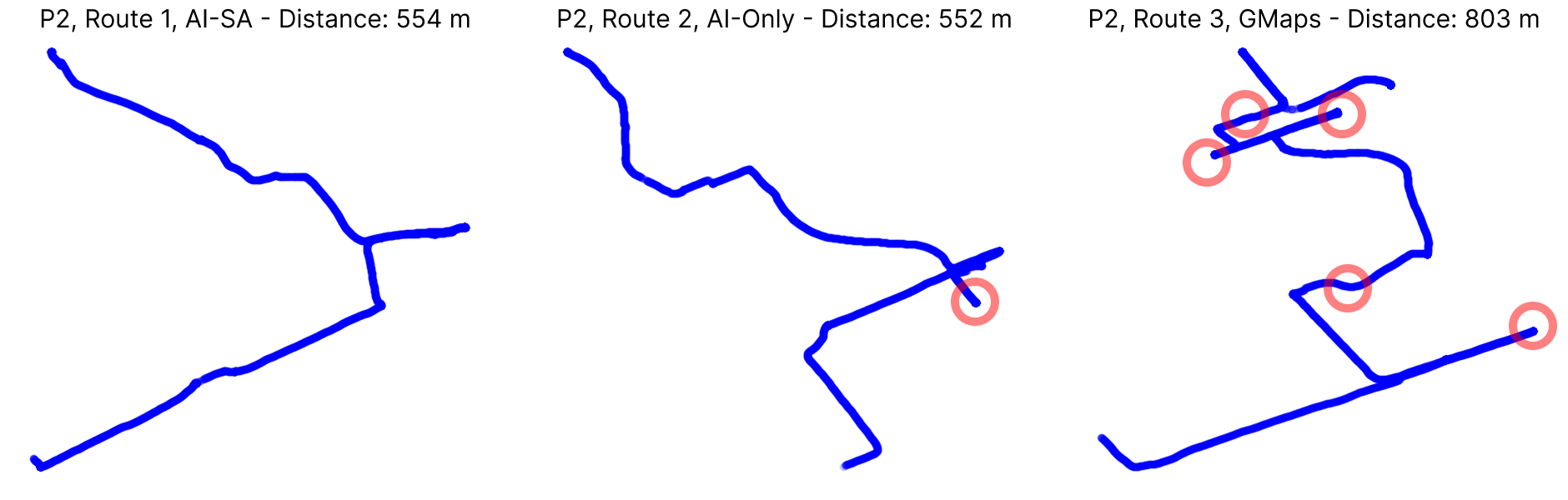}
    \caption{P2's walked paths with 0 deviations in Route 1, 1 in Route 2, and 5 in Route 3.}
    \label{fig:appendix-p2-paths}
\end{figure*}

\begin{figure*}[ht]
    \centering
    \includegraphics[width=.9\textwidth]{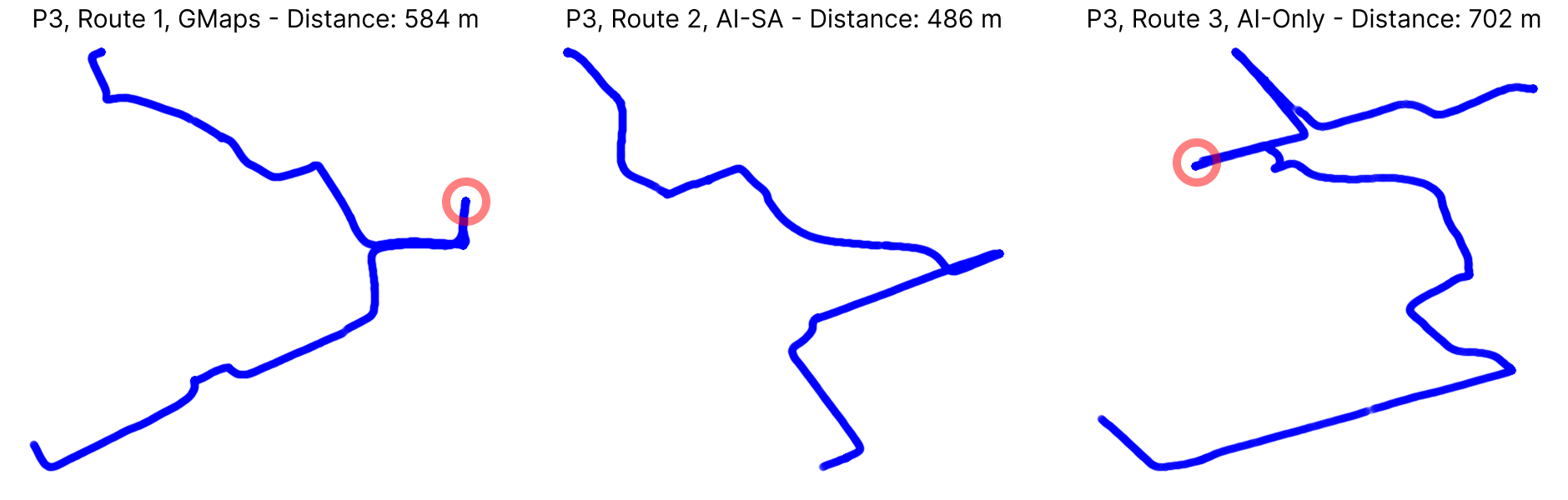}
    \caption{P3's walked paths with 1 deviation in Route 1, 0 in Route 2, and 1 in Route 3.}
    \label{fig:appendix-p3-paths}
\end{figure*}

\begin{figure*}[ht]
    \centering
    \includegraphics[width=\textwidth]{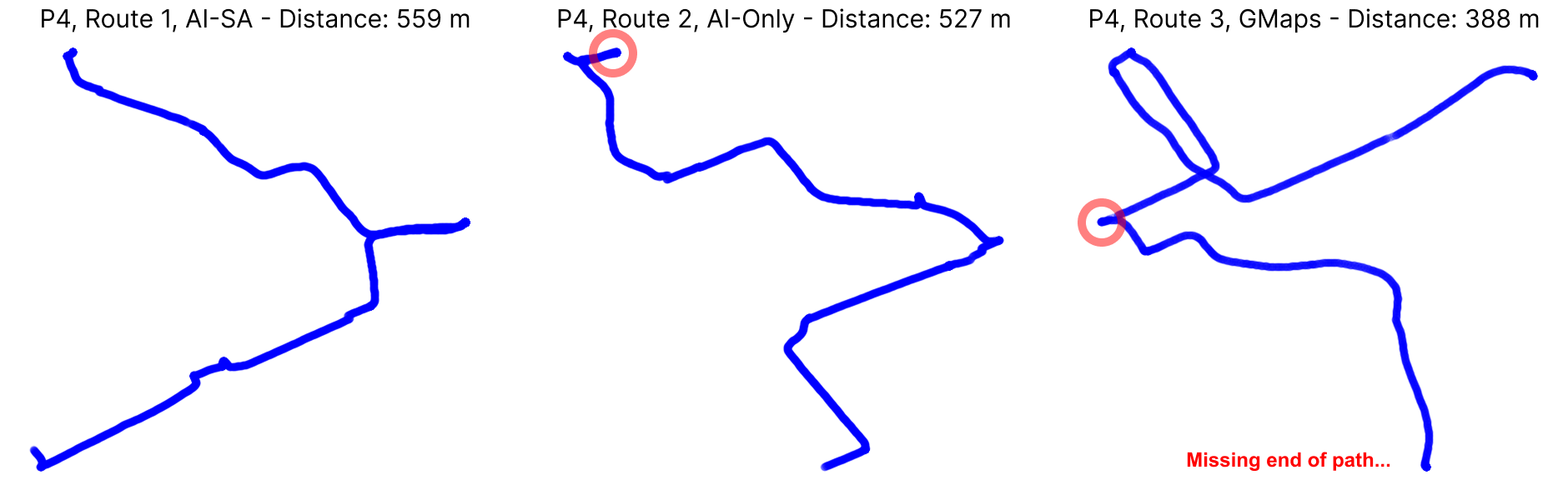}
    \caption{P4's walked paths with 0 deviations in Route 1, 1 in Route 2, and 1 in Route 3. Note that the last part of P4's GPS data is missing.}
    \label{fig:appendix-p4-paths}
\end{figure*}

\begin{figure*}[ht]
    \centering
    \includegraphics[width=\textwidth]{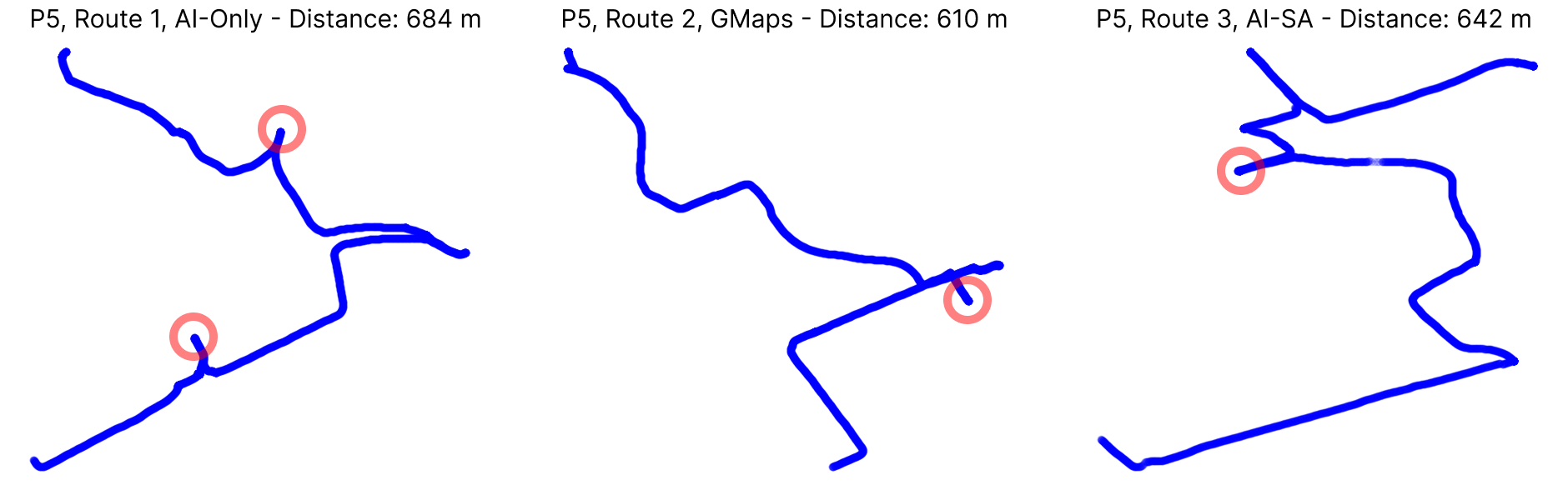}
    \caption{P5's walked paths with 2 deviations in Route 1, 1 in Route 2, and 1 in Route 3.}
    \label{fig:appendix-p5-paths}
\end{figure*}

\begin{figure*}[ht]
    \centering
    \includegraphics[width=\textwidth]{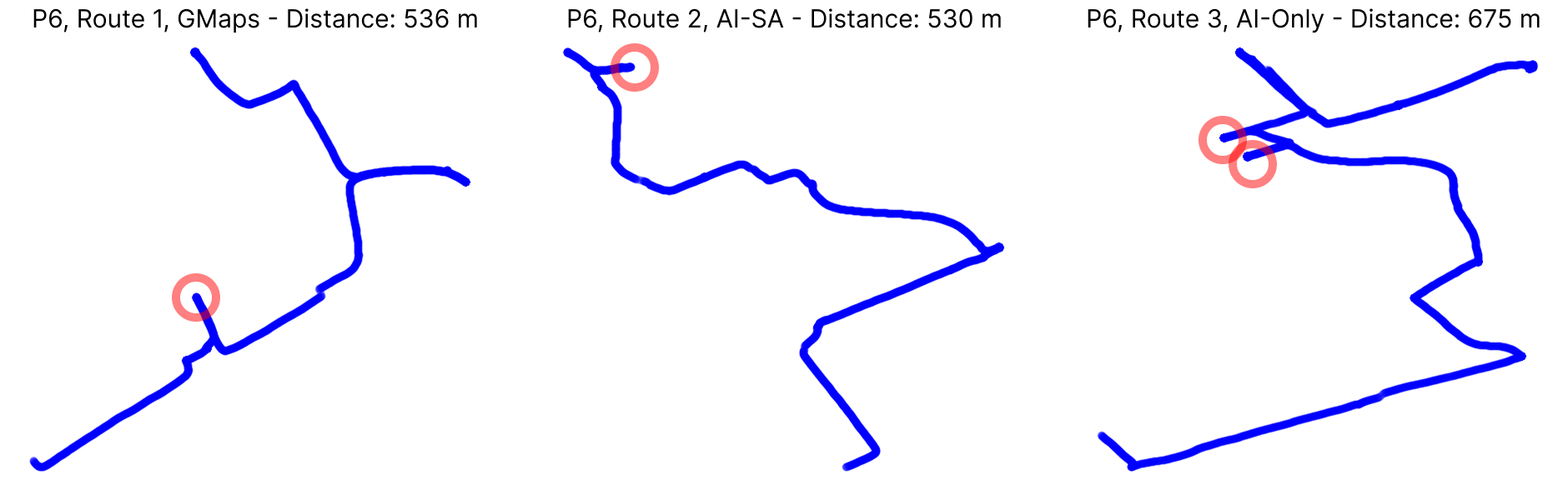}
    \caption{P6's walked paths with 1 deviation in Route 1, 1 in Route 2, and 2 in Route 3.}
    \label{fig:appendix-p6-paths}
\end{figure*}

\begin{figure*}[ht]
    \centering
    \includegraphics[width=\textwidth]{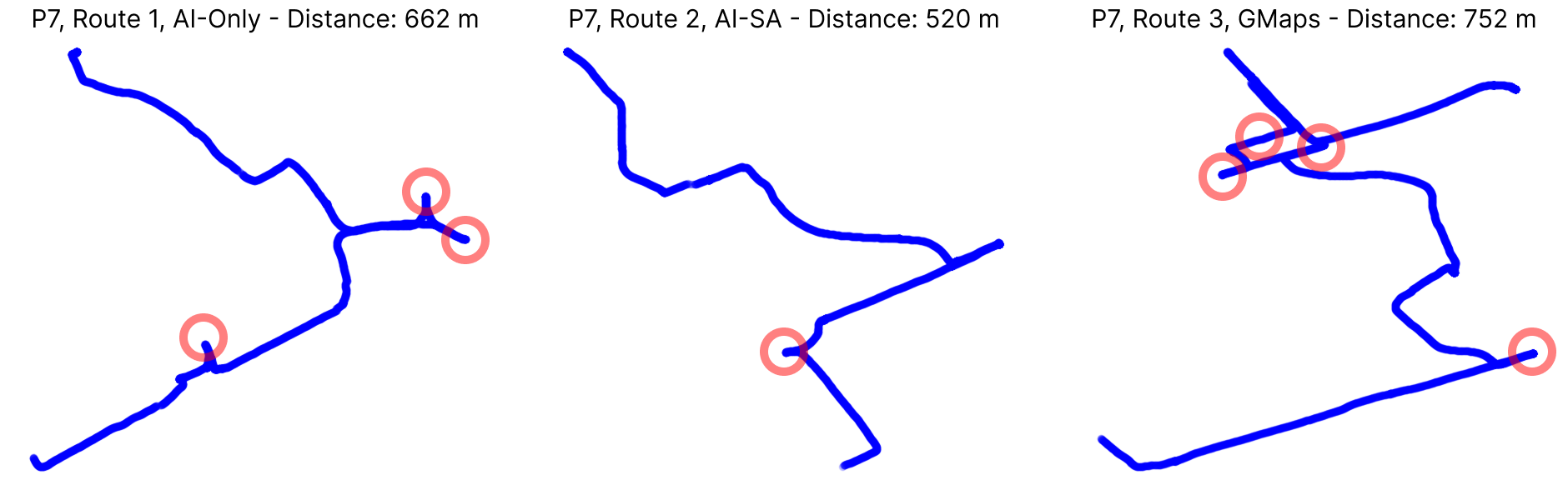}
    \caption{P7's walked paths with 3 deviations in Route 1, 1 in Route 2, and 4 in Route 3.}
    \label{fig:appendix-p7-paths}
\end{figure*}

\begin{figure*}[ht]
    \centering
    \includegraphics[width=\textwidth]{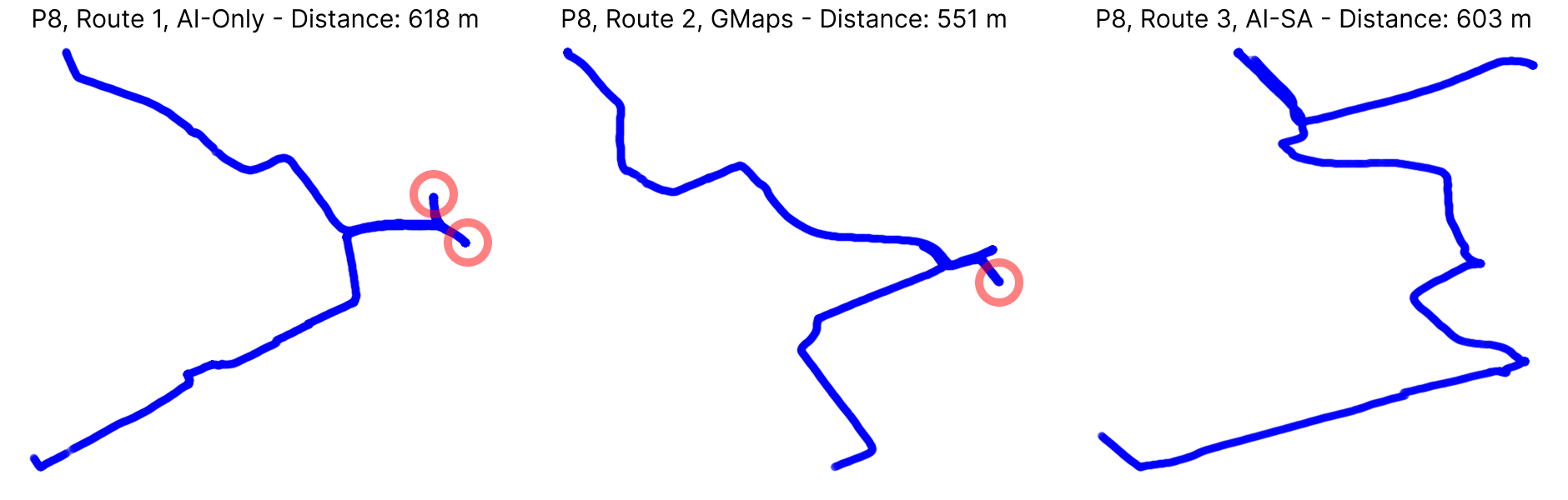}
    \caption{P8's walked paths with 2 deviations in Route 1, 1 in Route 2, and 0 in Route 3.}
    \label{fig:appendix-p8-paths}
\end{figure*}

\begin{figure*}[ht]
    \centering
    \includegraphics[width=\textwidth]{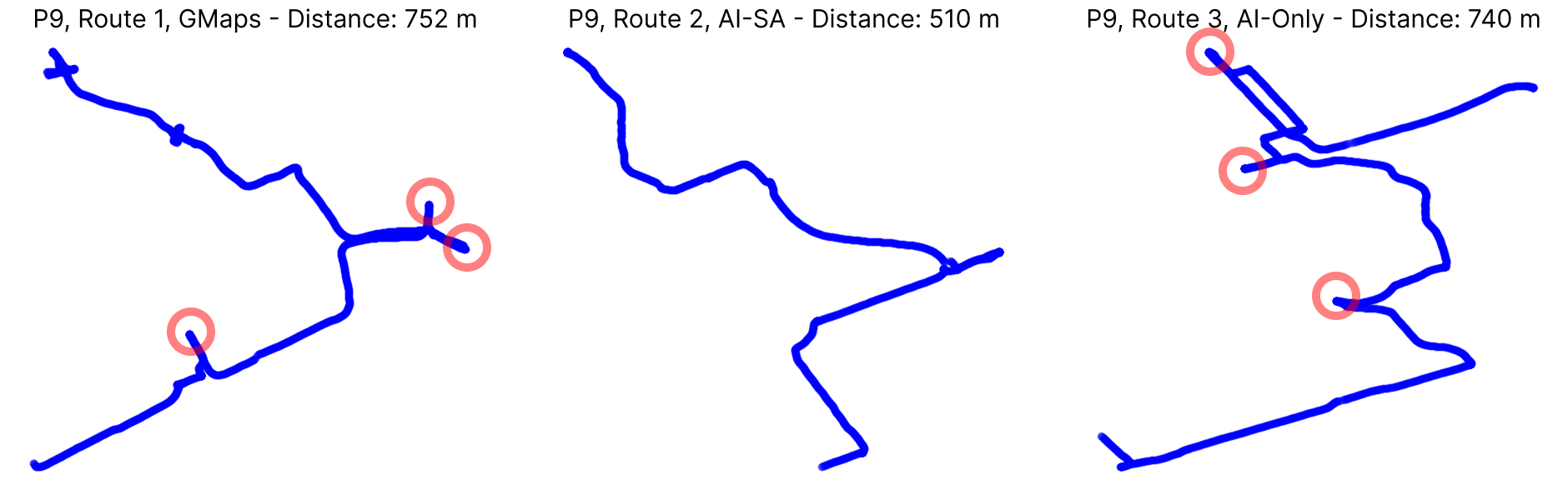}
    \caption{P9's walked paths with 3 deviations in Route 1, 0 in Route 2, and 3 in Route 3.}
    \label{fig:appendix-p9-paths}
\end{figure*}

\begin{figure*}[ht]
    \centering
    \includegraphics[width=\textwidth]{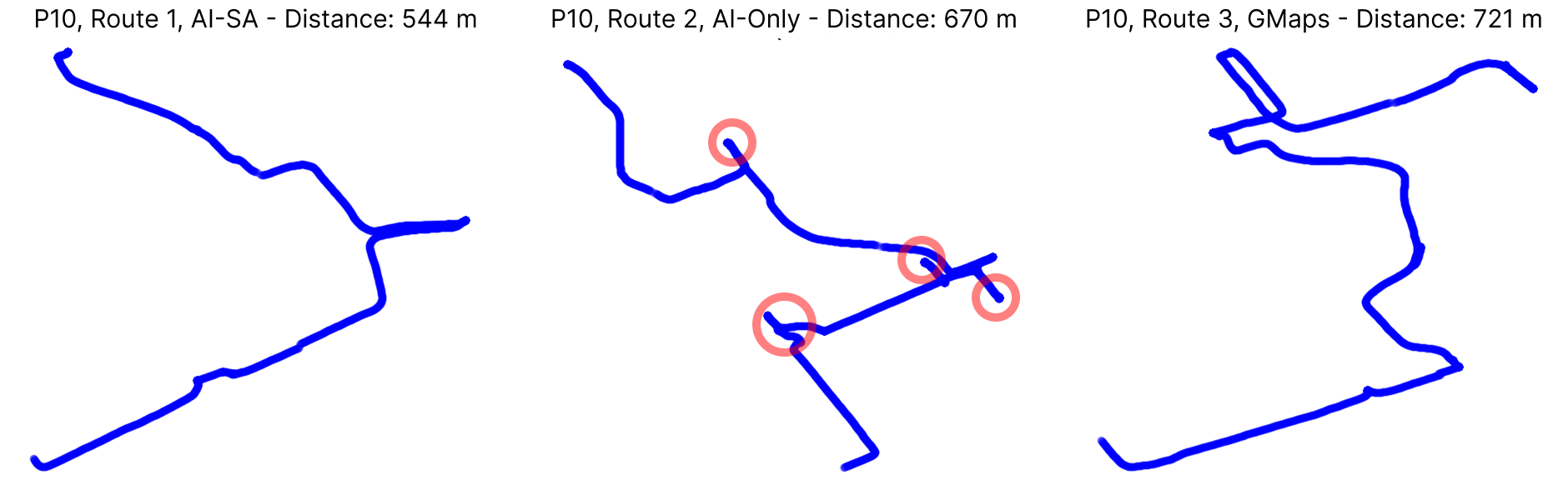}
    \caption{P10's walked paths with 0 deviations in Route 1, 4 in Route 2, and 0 in Route 3.}
    \label{fig:appendix-p10-paths}
\end{figure*}

\begin{figure*}[ht]
    \centering
    \includegraphics[width=\textwidth]{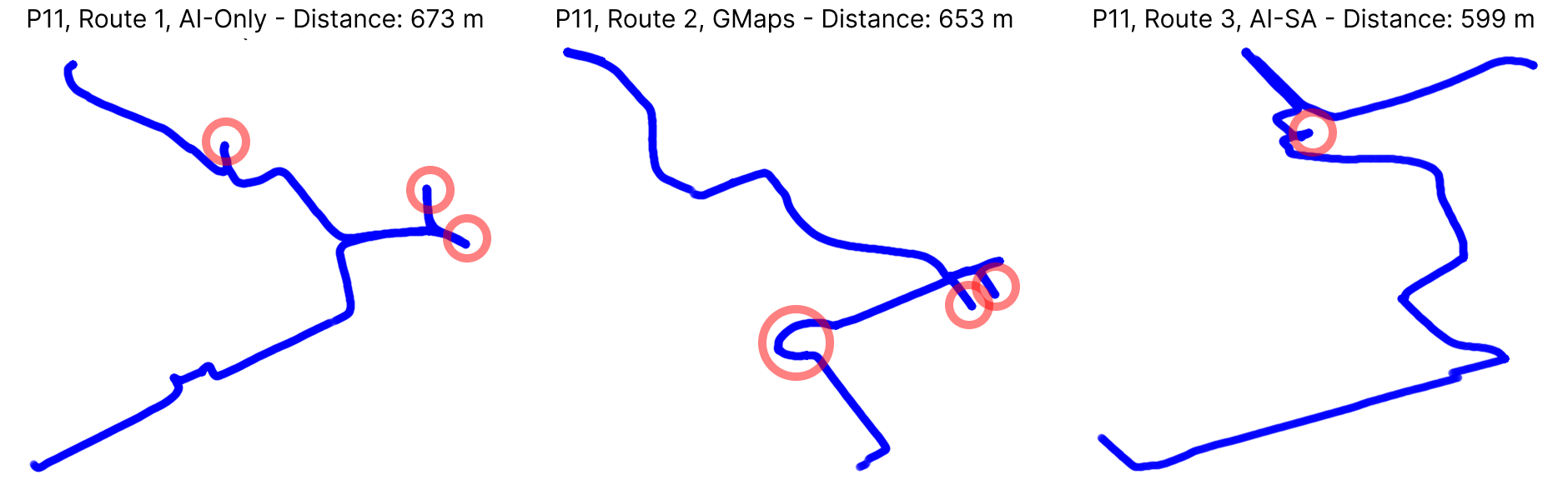}
    \caption{P11's walked paths with 3 deviations in Route 1, 3 in Route 2, and 1 in Route 3.}
    \label{fig:appendix-p11-paths}
\end{figure*}

\begin{figure*}[ht]
    \centering
    \includegraphics[width=\textwidth]{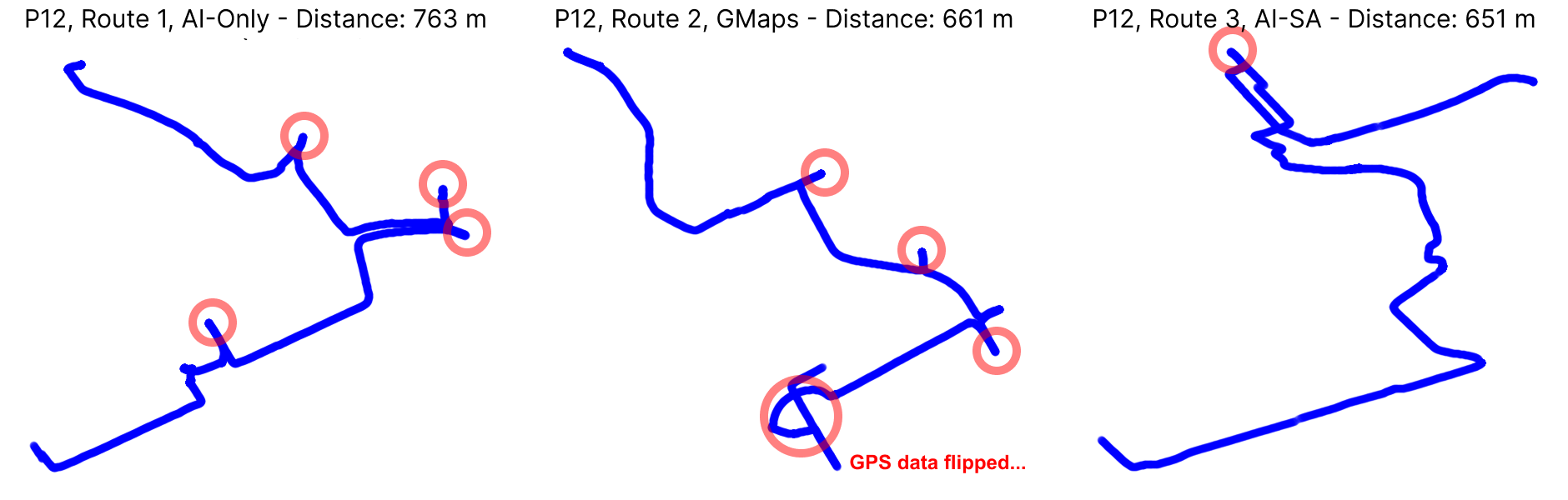}
    \caption{P12's walked paths with 4 deviations in Route 1, 4 in Route 2, and 1 in Route 3. Note that at the end of the path, the GPS/VPS data flipped; P12 continued down the correct path, as other participants. It is therefore not marked as a deviation.}
    \label{fig:appendix-p12-paths}
\end{figure*}

\end{document}